\documentclass[amsmath,amssymb,amsthm,aps,pra,twocolumn]{revtex4-2}
\usepackage[T1]{fontenc}
\usepackage{amssymb,amsmath}
\usepackage{amsfonts}

\usepackage{graphicx}
\usepackage{amsthm}
\usepackage{color}
\usepackage{enumerate}

\usepackage{graphicx}
\usepackage{dcolumn}
\usepackage{bm}

\usepackage{physics}
\usepackage{color}
\usepackage{hyperref}

\usepackage[smalltableaux, centertableaux, nobaseline]{ytableau}

\newcommand{\rd}[1]{{\color{red}#1}}

\newcommand{\wm}{{\color{white}-}}

\newtheorem{definition}[]{Definition}
\newtheorem{proposition}[]{Proposition}

\begin{document}

\preprint{APS/123-QED}

\title{
	Perfect quantum-state synchronization
}

\author{Jakub Czartowski$^1$, Ronny M{\"u}ller$^2$, Karol
  {\.Z}yczkowski$^{1,3}$, Daniel Braun$^2$}
 \email{daniel.braun@uni-tuebingen.de}
\affiliation{
$^1$Institute of Theoretical Physics,
Jagiellonian University,
ul. {\L}ojasiewicza 11, 30-348 Krak{\'o}w, Poland\\
$^2$Institute of Theoretical Physics,	University of T{\"u}bingen,
Auf der Morgenstelle 14, 72076 T{\"u}bingen, Germany \\
$^3$Center for Theoretical Physics, Polish Academy of Sciences, 
 02-668 Warszawa, Poland}		

\date{June 21, 2021}

\begin{abstract} 
We investigate the most general mechanisms that lead to perfect synchronization of the quantum states of all 
subsystems of an open quantum system starting from  an arbitrary initial state. We provide a necessary and sufficient condition for such ``quantum-state synchronization'', prove tight lower bounds on the dimension of the ancilla's Hilbert space in two main classes of quantum-state synchronizers, and
give an analytical solution for their construction. 
The functioning of the
found quantum-state synchronizer of two qubits is demonstrated experimentally on an IBM quantum computer and we show that the remaining asynchronicity is a sensitive measure of the quantum computer's imperfection.  
\end{abstract}

\maketitle

\section{Introduction}
Some of the most spectacular and technologically important quantum effects appear when a large 
number of quantum systems are in the same quantum state: Bose-Einstein condensation (BEC), superconductivity,
quantum magnetism, macroscopic quantum
tunneling,
or arguably even
lasing, where a large number of atoms emmit photons phase-coherently
into a mode of an electro-magnetic resonator.
In a superconductor or BEC a macroscopically occupied mode with a well
defined phase arises that manifests itself in macroscopic quantum
interference
relevant for instance in
superconducting quantum interference devices
(SQUID) for precision measurements of the magnetic field.  Particularly relevant for these macroscopic 
manifestations of quantum coherence is the synchronization of a
quantum phase, which does not have a classical counterpart. 
Such
``quantum-phase synchronization''
is therefore beyond the much
studied synchronization of quantum
dynamics of systems that synchronize classically \cite{PRK2001,walter_quantum_2014,walter_quantum_2015,DeVille2018SynchronizationAS}, 
such as coupled 
harmonic or non-linear oscillators \cite{heinrich_collective_2011,giorgi_quantum_2012,manzano_synchronization_2013,ludwig_quantum_2013,ludwig_quantum_2013,PhysRevLett.120.163601,PhysRevResearch.1.033012}.  
Synchronization of spins or spins and an oscillator was studied 
in \cite{giorgi_spontaneous_2013,zhirov_synchronization_2008,zhirov_quantum_2009,cattaneo_synchronization_2020,PhysRevResearch.2.023026,roulet_quantum_2018}.
Quantum-phase synchronization for two
qubits was examined in \cite{fiderer_quantum-phase_2016}, and for three-level systems in \cite{Jaseem2020QuantumSI}.

The goal of this work is to
 investigate the most general quantum channels that synchronize not only relevant quantum phases but 
the {\em full
quantum states of all subsystems}, understood as the reduced states of
the many-body system. 
We dub this kind of synchronization 
``Quantum-State 
Synchronization'' (QSS). We hence turn around the so far prevailing approach
of studying quantum synchronization for given systems
and ask {\em what are the most general quantum channels that lead to 
  synchronized behavior of quantum systems?}
The hope is that this approach will ultimately lead to engineering new macroscopic quantum effects.   \\

\section{Setting the scene}

Consider   a composite system {with } Hilbert space
$\mathcal{H}=\mathcal{H}_A\otimes \mathcal{H}_S$ with a ``system'' Hilbert space that is itself composed of $n$ subsystems, $\mathcal{H}_S=\mathcal{H}_1 \otimes \hdots \otimes \mathcal{H}_n$.  The 
dimension $N=\text{dim}\mathcal{H}_i$, $i=1,\ldots,n$ is the same for all $n$
subsystems, whereas the ancillary Hilbert space $\mathcal{H}_A$ might have different dimension $M$. 
Let $\mathcal B(\mathcal H_i)$ be the set of bounded linear hermitian operators
on $\mathcal H_i$. A quantum channel $\Phi: \mathcal B(\mathcal H_S)\to
B(\mathcal H_S)$, $\rho\mapsto\rho'= \Phi(\rho)$  is a linear,
trace-preserving completely positive map. A partial trace that
returns the reduced state of the 
$i$-th subsystem will be 
denoted as $\Tr_{\bar{i}} \rho$, where $\bar{i}$ stands for 
the complement of the $i$-th subsystem.
\begin{definition}
  We call a   state
  $\rho_\text{sync}\in \mathcal B(\mathcal H_S)$ a
``synchronized quantum state'' (SQS) if  all its reductions to 
single-party systems are identical, 
	\begin{equation}
		\Tr_{\bar{i}} \rho_{sync} = 
		\Tr_{\bar{j}} \rho_{sync}\label{eq.def}
	\end{equation}
	for all $i,j\in\qty{1,\hdots,n}$. A quantum channel $\Phi$ is
        called a ``quantum-state 
synchronizer'' (QSSR) 
if and only if for arbitrary initial $\rho\in \mathcal
B(\mathcal H_S)$, $\rho'=\Phi(\rho)$ is
a quantum-synchronized state.
\end{definition}
The definition leads to three immediate consequences: 1.) The set $\cal
S$ of all QSSRs for given $n,N$ is closed under concatenation, i.e.~with
$\Phi_1,\Phi_2\in \cal S$, also $\Phi_2\circ\Phi_1\in \cal S$.
2.) $\cal S$ is a convex set, i.e.~$\forall\,\Phi_1,\Phi_2\in \cal S$, also
$p\Phi_1+(1-p)\Phi_2\in {\cal S}\, \forall\, p\in [0,1]$.
3.) We can restrict the study of such channels to pure input states. This
follows from linearity of a channel and the partial trace, and the $P$-function
representation of any state -- see Appendix. \\
One trivial example of a SQS is the maximally mixed
state, $\mathbb{I}/N^n$, which has all reductions identical to the
local maximally mixed state $\mathbb{I}/N$. Another class of such states are 
maximally entangled 
 states for which reduction to any 
subsystem results in a maximally mixed state. Even more general SQSs are permutationally invariant states.
These are states for which $P_{ij}\rho P_{ij}=\rho$, where
$P_{ij}$ permutes subsystems $i,j$. 
Correspondingly, a quantum
channel that describes relaxation to thermal equilibrium at infinite
temperature is a special (albeit trivial) case of a QSSR, and so is a
quantum channel that resets all input states to a maximally entangled
state. 
While illustrating opposite extremes
of synchronizing channels' spectrum, the mentioned examples are not so interesting, 
as all quantum coherences are destroyed. In
general, one would like to get not only QSS, but also keep the synchronized states as
pure as possible, and achieve the synchronization with as small an
ancilla as possible.  

QSS is distinct from quantum cloning, where an arbitrary unknown state $\ket{\psi}$
is ``copied'' onto a fixed blank state $\ket{0}$, $\ket{\psi}\otimes\ket{0}\mapsto
\ket{\psi}\otimes\ket{\psi}$, which is possible only approximately \cite{Wootters82,GisinM97}. Note that  QSS is not excluded by the quantum
no-broadcasting theorem \cite{Barnum96}.
In analogy to selfcomplementary channels \cite{SRZ16,czartowski_trade-off_2019} 
 QSS maps arbitrary initial  
states to identical reduced states that are not necessarily
the same as any of the initial ones nor is the total final state required
to be a product state.  
\\

\section{Necessary and sufficient conditions for SQSs}

The pure state $\ket{\psi}\in \mathcal H_S$ can be decomposed into
a linear superposition of 
states that are either even or odd under permutation of two subsystems
$\ket{\psi}=(\ket{\psi}+P_{ij}\ket{\psi})/2+(\ket{\psi}-P_{ij}\ket{\psi})/2\equiv
\ket{s_{ij}}+\ket{a_{ij}}$,
where we use $P_{ij}^2=\mathbb{I}$, and $i,j=1,\ldots,n$ (see Appendix for possible generalizations for ``anyonic'' states).
We are only concerned with permutations of the system's subsystems, not the ancilla.  The states
$\ket{a_{ij}},\ket{s_{ij}}$ are not normalized in general.  In particular, if $\ket{\psi}$ is (anti-)symmetric, $\ket{a_{ij}}$ ($\ket{s_{ij}}$) vanishes, respectively.  Under partial
trace we find $\forall j$ 
\begin{eqnarray}
  \label{eq:ptr}
\rho_i  &=&\Tr_{\bar
      i}\qty(\ketbra{a_{ij}}+\ketbra{s_{ij}})\nonumber\\
  &&+\Tr_{\bar
      i}\qty(\dyad{a_{ij}}{s_{ij}}+\dyad{s_{ij}}{a_{ij}})\,.\label{eq:ptr2}
\end{eqnarray}
The reduced state
$ \rho_j$ can be found by first permuting  
systems $i,j$, $\rho_j=\Tr_{\bar i} P_{ij} (| \psi \rangle \langle \psi | P_{ij}) \,\,\,\forall i $.  Using the
transformation properties of states $\ket{a_{ij}},\ket{s_{ij}}$, we see that 
$\rho_j$ differs from $\rho_i$ only by the sign of the second partial trace in
eq.\eqref{eq:ptr2}. This implies the following condition for SQSs:
\begin{proposition} \label{prop:1}
A
state $\ket{\psi}\in \mathcal{H}_S$ is a SQS iff $\Tr_{\bar
      i}(\ketbra{a_{ij}}{s_{ij}}+\ketbra{s_{ij}}{a_{ij}})=0$ for all $i,j\in\{1,\ldots,n\}$\,. \label{cond.gen}
  \end{proposition}
This condition is easily extended to a state in the
full Hilbert space, $\ket{\psi}\in
\mathcal H$ that serves as purification of a mixed state
$\rho\in\mathcal{B}(\mathcal{H}_S)$ by simply including  the ancilla in $\bar i$.
The condition in Proposition~\ref{cond.gen} is equivalent to requiring that $\Tr_{\bar
      i}|a_{ij} \rangle \langle s_{ij} | $ is purely anti-hermitian for all $i,j$ (this expression is still an operator on system $i$).  Several sufficient conditions follow: A state is a SQS,
$1.)$ if $\Tr_{\bar i}| a_{ij}\rangle \langle s_{ij}|=0$ $\forall\, i,j\in\{1,\ldots,n\}$;
or $2.)$ if $\ket{\psi}$ has definite symmetry or anti-symmetry 
under $P_{ij}$, i.e.~$\ket{a_{ij}}=0$ or $\ket{s_{ij}}=0$ for all $i,j$.  In this case
$\dyad{\psi}$ is a permutationally symmetric state, but we see that
the set of SQSs is larger than the set of permutationally symmetric
states.  An example of such a state that is not permutationally
symmetric is the two-qubit state $\ket{\psi}=(0,1,i,0)/\sqrt{2}$
written in computational basis. It leads to
$\rho_1=\rho_2=\mathbb{I}/2$, i.e.~$\ket{\psi}$ is a (trivial) SQS, but the full density matrix $\ketbra{\psi}{\psi}$ is not permutationally symmetric. More involved examples include absolutely maximally entangled (AME) states for the systems of five and six qubits, AME(5,2) and AME(6,2) respectively, for which $\forall_i \rho_i = \mathbb{I}/2$. Hence these states are synchronized, despite lack of symmetry or antisymmetry.
We prove in Appendix that one cannot mix different
symmetries for different pairs $i,j$ in condition 2.). 

\section{Minimal ancilla}

First focus on condition 2.) with full
permutational symmetry of the output state. 
We then need to
find transformations that map any pure input 
state to fully symmetrized states. A synchronizing channel that achieves this will be
called a ``symmetrizing QSSR''. 
        To this end we represent the
        quantum channel $\Phi$ without restriction of generality as 
	\begin{equation}
		\Phi(\rho) = \Tr_E\qty[U\qty(\op{0}_E\otimes\rho)U^\dag],
	\end{equation}
        where $\ketbra{0}_E$         is a fixed state of an
        ancilla $E$, and $U$ a joint 
        unitary evolution of the system of the $n$ 
        qudits and the ancilla. Clearly, to achieve a fully
        symmetric output state for all pure input states, it is sufficient and necessary
        to do so for all computational basis states, i.e.~when
        including the initial state of the ancilla, all states of
        the form $\ket{0;j_1,\ldots,j_n}$, $j_k=1,\ldots,N$. 
This state is mapped to a column of 
$U$ with components
$\bra{k;i_1,\ldots,i_n}U\ket{0;j_1,\ldots,j_n}=U_{k;i_1\ldots i_n,0;j_1\ldots j_n}$.
Such a column then needs to be a vector of the form
\begin{equation}
  \ket{\psi_\text{sym}} = \sum_{k=1}^M
  \sum_{i_1,\ldots,i_n} a_{k;\qty{i_1,\hdots,i_n}}\ket{k;i_1,\hdots,i_n},
	\end{equation}
	where the expansion coefficients $a_{k;\qty{i_1,\hdots,i_n}}$
        depend on an unordered set of 
        indices of $S$. 
        We 
        denote the space spanned by such states as
        $\mathcal{H}_\text{sym}^{N,n} =        \text{span}\qty(\ket{\psi_\text{sym}})$.
These states span a subspace of dimension 
	\begin{equation}
\text{dim}(\mathcal{H}_\text{sym}^{N,n})=		M \binom{n+N-1}{n}\,.\label{dimHsym}
	\end{equation}
Since the $N^n$ computational basis states (extended by the fixed
initial state of the ancilla) map to the first $N^n$ columns of
$U$, and these need to be orthonormal, the symmetric subspace
must accomodate at least $N^n$ 
linearly independent images of the computational basis states.
This sets a tight lower bound on the dimension $M$ of the ancillary
Hilbert space. We have thus
proved the following necessary and sufficient 
condition for the existence of a symmetrizing QSSR:
\begin{proposition}
The tight lower bound on the dimension $M$ of the ancilla necessary to perfectly
quantum-state synchronize with a symmetrizing QSSR $n$ 
qudits of dimension $N$ prepared in an
arbitrary initial state, is 
	\begin{equation}
		M \geq N^n \binom{n+N-1}{n}^{-1} \overset{N\rightarrow\infty}{\longrightarrow}n!\,.
	\end{equation}
      \end{proposition}
For $n=2$, $M=2$ is the tight lower bound for any $N$, i.e.~two qudits of arbitrary dimension can be synchronized with a single qubit as ancilla. 
And a single harmonic oscillator as ancilla can lead to perfect QSS for any number of qudits. A short table giving minimal dimensions for low $n$ and $N$ is given in 
Appendix.
Our result does not contradict earlier findings \cite{PhysRevLett.121.053601} according to which the smallest quantum system that can synchronize is a spin-1, as that work uses a different definition of synchronization, requiring the existence of a limit cycle.

Alternatively,
the channel $\Phi$ can be represented
in terms of $M$ Kraus operators $K_a$ satisfying the identity resolution condition, $\sum_{a=1}^M K_a^\dag K_a = \mathbb{I}$,
	\begin{equation}
		\Phi(\rho) = \sum_{a=1}^M K_a \rho K_a^\dag. 
	\end{equation}
The first $N^n$ columns of $U$
define the $M$ Kraus operators, which can be read off as $N^n\times
N^n$ blocks stacked in those first $N^n$ columns of $U$ by the reshuffling $(K_a)_{ij}=U_{(a-1)N^n+i,j}, \ i,j= 0,...,N^n-1$.  Since an
entire column vector of $U$ has definite permutational symmetry, the
columns of all Kraus operators must have the same symmetry.\\ 

\section{Construction of quantum-state synchronizers}

\begin{figure}[h!]
	\begin{center}
		\includegraphics[width=\columnwidth]{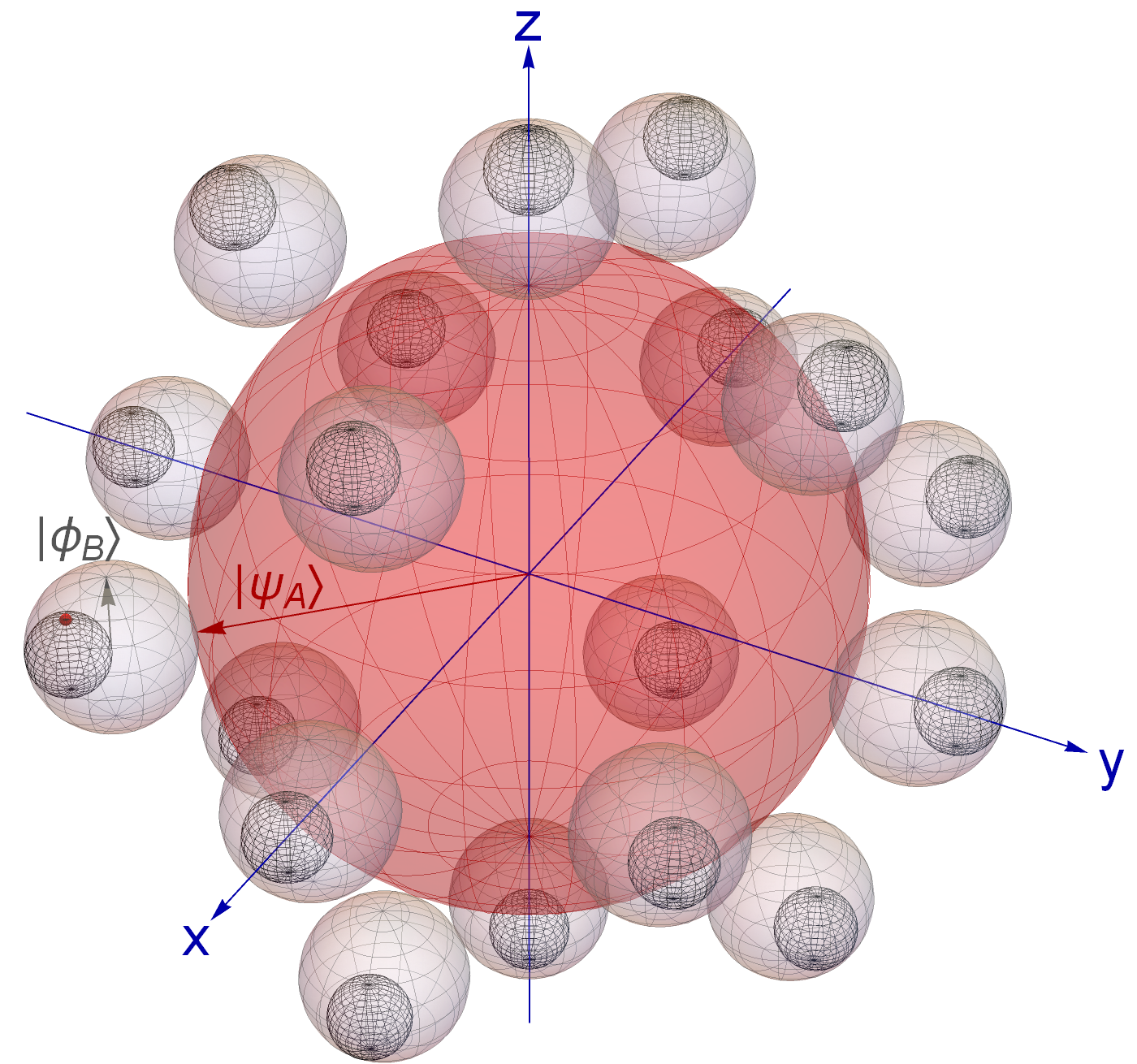} 
		\caption{Action of the channel with Kraus operators \eqref{kraus} on product states, $\ket{\psi} = \ket{\psi_A}\otimes\ket{\phi_B}$. The positions of gray spheres on the central red sphere represent $\ket{\psi_A}$. The gray spheres are Bob's Bloch ball, i.e.~the position of a point on the gray spheres the initial state of Bob $\ket{\phi_B}$. The small black spheres depict the end-state of the QSSR, $\rho_A = \rho_B = \Tr_B(\Phi(\op{\psi}))$. 
                  The grids on the small black spheres are exactly aligned with the gray spheres which allows one to 
                  identify 
                  the resulting state based on the positions on the red and gray spheres, 
                  i.e.~$\rho_B$ corresponds to the state on the small black sphere (given through its Bloch vector with respect to the gray sphere) with coordinates of $\ket{\phi_B}$ on the gray sphere. 
                  The entire figure is rotationally symmetric about any axis through the origin. }\label{ball2}
	\end{center}
      \end{figure}

Next, we 
        provide the construction for a symmetrising QSSR. We illustrate
         each step by the simplest possible example of
        two qubits, $n = N = 2$, with minimal ancilla $M = 2$. 
	First, 
        choose an arbitrary unitary matrix $U$ of dimension $M
        \binom{n+N-1}{n} = 6$ and 
        select the first $N^n = 4$ columns from it. For instance~
	\begin{equation}
		U = \mqty(
				1 & 0 & 0 & 0 & 0 & 0 \\
				0 & \frac{1}{\sqrt{2}} & \frac{1}{\sqrt{2}}& 0 & 0 & 0 \\
				0 & 0 & 0 & 1 & 0 & 0 \\
				0 & 0 & 0 & 0 & 1 & 0 \\
				0 & \frac{1}{\sqrt{2}} & \frac{-1}{\sqrt{2}} & 0 & 0 & 0 \\
				0 & 0 & 0 & 0 & 0 & 1
			)
			\rightarrow
			\mqty(
				1 & 0 & 0 & 0 \\
				0 & \frac{1}{\sqrt{2}} & \frac{1}{\sqrt{2}} & 0 \\
				0 & 0 & 0 & 1 \\\hline
				0 & 0 & 0 & 0 \\
				0 & \frac{1}{\sqrt{2}} & \frac{-1}{\sqrt{2}} & 0 \\
				0 & 0 & 0 & 0
			),\label{11}
	\end{equation}
          written in the basis states $\ket{j;i_1,\hdots,i_n}_{S}$
of the $M \binom{n+N-1}{n}$-dimensional space, 
		 	where $0 \leq i_1 \leq \hdots \leq i_n \leq N-1$ and $j \in
        \qty{1,\hdots,M}$. 
        Now embed 
        the $\ket{j;i_1,\hdots,i_n}_{S}$ in 
        the $MN^n$-dimensional space as
	\begin{equation}
		\ket{j;i_1,\hdots,i_n}_S \rightarrow \frac{1}{\mathcal{N}}\sum_{\sigma\in S_n} \ket{j;\sigma(i_1,\hdots,i_n)}\,,\label{9}
	\end{equation}
	with $S_n$ the permutation group on $n$ elements and $\mathcal{N}$ the normalisation factor introduced in order to keep unit norm for all basis states.
For $n=2$, the matrix elements that define the map \eqref{9}
are SU(N) Clebsch-Gordan coefficients. The symmetric and antisymmetric cases correspond to the highest and lowest ``total angular momentum''.
  E.g.~for $n=2$, an $N$ state system is a pseudo-spin $(N-1)/2$.  In the fully symmetric case, $$\ket{j;i_1,i_2}\mapsto \ket{j}\sum_{M=-(N-1)}^{N-1}
  \ket{J,M}\braket{J,M}{j_1,j_2;m_1,m_2}$$
  with $J={N-1}$, $m_k=i_k-j_k$, $j_k=\frac{N-1}{2}$, $k=1,2$, in standard $j,m$-notation of angular momenta. For $n=3$, the matrix elements are $3j$-symbols, and so on.  For arbitrary $n$ they can always be calculated by applying the descending collective angular momentum ladder operator to the state with $i_k=N-1$ $\forall k$  (see Appendix for details). Classically, this procedure is inefficient in $n$ due to the exponential increase of the number of computational basis states for which matrix elements are needed, but an efficient quantum algorithm for the calculation of Clebsch Gordan coefficients exists \cite{PhysRevLett.97.170502,u_efficient_2013}. 
 In the
          2-qubit case
$		\ket{j;00} \rightarrow \ket{j;00}$, 
$	\ket{j;01}  \rightarrow \frac{1}{\sqrt{2}}\qty(\ket{j;01} + \ket{j;10})$, 
$		\ket{j;11}  \rightarrow \ket{j;11}$.
After the embedding, 
revert from the description with
        $MN^n$-dimensional column vectors of $U$  to $M$ Kraus 
        operators $K_i$,  
	\begin{align}\label{kraus}
		K_1 = \mqty(
			1 & 0 & 0 & 0 \\
			0 & \frac{1}{2} & \frac{1}{2} & 0 \\
			0 & \frac{1}{2} & \frac{1}{2} & 0 \\
			0 & 0 & 0 & 1)\,,
		&&
		K_2 = \mqty(
			0 & 0 & \wm0 & 0 \\
			0 & \frac{1}{2} & -\frac{1}{2} & 0 \\
			0 & \frac{1}{2} & -\frac{1}{2} & 0 \\
			0 & 0 & \wm0 & 0
		)\,.
	\end{align}
Fig.~\ref{ball2} summarizes the action of this channel. Remarkable is the rotational symmetry about any axis passing through the center of Alice's Bloch sphere, implying that already synchronized states are mapped to themselves. The channel is non-unital and leaves large quantum coherences.
For the second possibility in condition 2.), i.e. that
the first $N^n$ columns of $U$ are fully antisymmetric, 
the dimension is $M\binom{N}{n}$; the
rest of the argument is the same as in the fully symmetric
case. Hence, 
we have $M\geq N^n/\binom{N}{n}$, which is always larger than in the symmetric
case. \\
One way to obtain more general QSSRs 
is to exploit the condition in Proposition \ref{cond.gen}, without imposing permutational
symmetry in the columns of $U$.  A simple example is given by a
channel with the same Kraus operators as in eq.\eqref{kraus} up to an
overall minus sign in the third 
row in $K_2$.
The channel maps any state to a mixture of symmetric and
anti-symmetric states, hence a permutationally symmetric state and
hence a SQS. Allowing $\pm$ symmetry independently in $M$ Kraus
operators, $M+1$ different types of QSSRs arise (only the number of (anti-)symmetric Kraus operators matters), where the largest $M$
to be considered is $N^{2n}$. Even more generally, the representation of $P_{ij}$ can be different between the Kraus operators. An example of such behaviour, which leads to a QSSR manifold with additional $M\qty(\frac{N(N-1)}{2}+1)\qty(n-1)$ dimensions, is given in Appendix.

Necessary conditions for the existence of more
general QSSRs can be obtained from parameter counting. A single
normalized pure states $\ket{\psi}\in \mathcal H$ contains $2(MN^n -
1)$ real parameters, not counting the global phase. Eq.\eqref{eq.def} are
$(n-1)(N^2-1)$ conditions. This leaves  $2(MN^n - 1)- (n-1)(N^2-1)$
free real parameters in $\ket{\psi}$ in order to have a SQS after
tracing out the ancilla. Even though the eqs.\eqref{eq.def} are
non-linear in the components of $\ket{\psi}$, this parameter count
sets a mild (for $n>2$) lower bound on $M$.\\

\section{Experimental realization on a quantum computer}

We
verified the functioning of the 2-qubit QSSR with a single-qubit ancilla experimentally by  
constructing a quantum circuit that realizes \eqref{kraus} -- see Appendix --
 and executing the quantum circuit 
on the five qubit ``Santiago'' quantum computer of IBM, a noisy
intermediate-scale quantum (NISQ) device. 

We evaluate the quality of synchronization after the action of the QSSR with an asynchronicity measure $\mathcal{A}$ based on the spread of the Bloch vector components $\hat{ x}^{(i)}_j$ of the individual
qubits, $i=1,\ldots,n$.
  They are estimated as empirical means from
$N_r$ measurements for each component, $\hat
x^{(i)}_j=(1/N_r)\sum_{\nu=1}^{N_r}s_j^{(i,\nu)}$, where $s_j^{(i,\nu)}\in \{\pm
1\}$ is the measurement outcome of measuring the Pauli matrix
$\sigma_j^{(i,\nu)}$ in the $\nu$-th run with a given initial state, and $j=1,2,3$ for $x,y,z$.
To prevent strongly mixed states from contributing
little inspite of large angular spread of the Bloch vectors, 
the Bloch vectors 
are rescaled so
that the longest Bloch vector has unit length, 
\begin{align*}
     \Tilde{x}^{(i)}_j &= \hat{x}^{(i)}_j / r_\text{max} \quad \quad r_\text{max} = \max \{r^{(1)},...,r^{(n)}\}\\
r^{(i)} & = \sqrt{(\hat{x}_1^{(i)})^2+(\hat{x}_2^{(i)})^2+(\hat{x}_3^{(i)})^2} \quad \Tilde{\mu}_j = \frac{1}{n} \sum_{i=1}^{n}
\Tilde{x}^{(i)}_j\, .
 \end{align*}
The measure $\mathcal{A}$  for a given initial
state is then defined as
\begin{equation}
\label{bench}
    \mathcal{A} = \sum_{j=1,2,3} \frac{1}{n} \sum_{i=1}^{n} (\Tilde{x}^{(i)}_j -\Tilde{\mu_j})^2\,.
\end{equation}

On perfect devices the estimates of the Bloch vectors
are still subject to statistical noise, which results in a finite value of
${\mathcal{A}}$ for finite $N_r$.  The noise 
 decays as $1/N_r$ for large $N_r$, such that asynchronicity   
$\mathcal{A}=0$ signals perfect QSS in that case.  For imperfect
QSS, to be expected on a NISQ device, $\mathcal{A}$ will remain finite even for
$N_r\to\infty$.  

\begin{figure}[h]
	\begin{center}
		\includegraphics[width=\columnwidth]{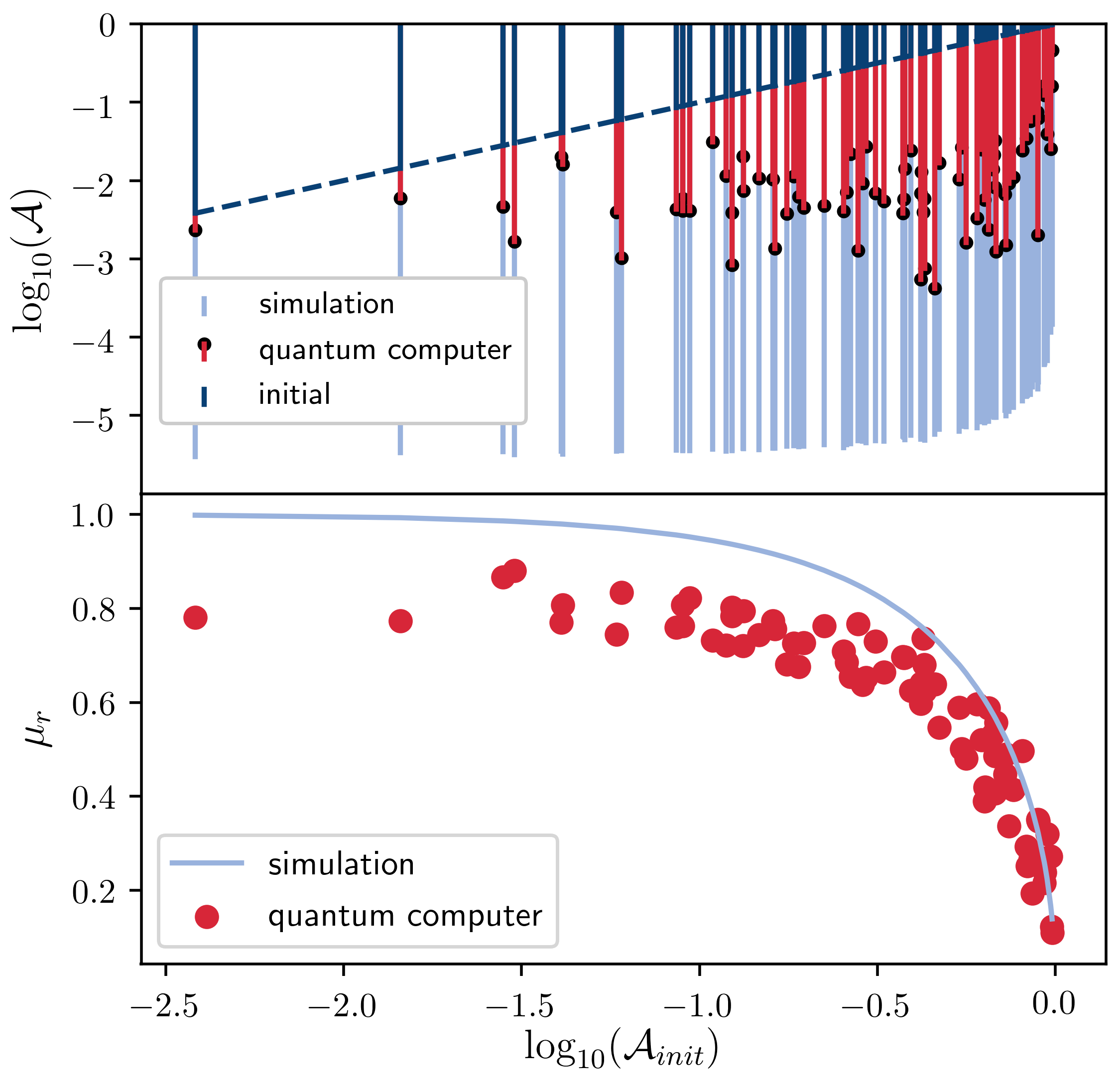} 
		\caption{\label{results}  
		Experimental implementation of the quantum channel \eqref{kraus}.
		Upper part: Each bar contains the asynchronicity of the initial state
                ($\mathcal{A}_{init}$,
                lower end of dark blue lines
                ), of the final experimental state ($\mathcal A$,
                black dots),
		simulated state (light blue), plotted against ${\mathcal{A}_{init}}$. The simulation
		data is free of hardware noise and shows the statistical limit set by the finite number of shots. 		  {The dashed blue line is given by $\mathcal A=\mathcal A_{init}$}. 
		Lower part: Experimental and simulated mean Bloch vector lengths
		$\mu_r = \frac{1}{n}\sum_{i=1}^{n}r^{(i)}$ as  function of the
		initial asynchronicity. 
	} 
	\end{center}
\end{figure}
We used $N_r=204800$ shots to estimate each Bloch
vector component of 
 $N_s=77$ random initial states.  The experimental
results of ${\mathcal{A}}$ and the mean Bloch vector lengths 
are shown  in Fig. \ref{results}.
Despite the imperfect hardware, the QSSR reduces the
asynchronicity of initial states by up to several orders of
magnitude. The drop in ${\mathcal{A}}$ is largest for initially not well synchronized states,
as the final value of ${\mathcal{A}}$ shows little variation.
Simulated results for perfect hardware are included to act as a
reference to the statistical limit, set by the finite number of shots.
The 
difference in simulation and experiment is due to the noisy hardware,
which is not accounted for in the simulation.  We see that the
asynchronicity measure is a much more sensitive measure of the
hardware errors than the reduction in purity of the qubits given by
the mean Bloch vector lengths. To assess the overall performance of the QSSR, we average
$\mathcal{A}$ over $N_s$ 
randomly chosen Haar distributed
initial states and define the mean asynchronicity 
$\overline{\mathcal{A}}=\frac{1}{N_s} \sum_{\gamma=1}^{N_s} \mathcal{A}_{\gamma}$,
where  $\gamma$ labels the initial states. 
In our experiments it  drops by more than an order of magnitude, 
$
    \text{initial: } \overline{\mathcal{A}}=0.44 \rightarrow \text{final: } \overline{\mathcal{A}}=0.027\,.
$

\section{Summary}

We have found necessary and sufficient conditions and an analytical construction of non-trivial 
QSSRs that perfectly synchronize the quantum
states of an arbitrarily large number of quantum systems in arbitrary initial
states.  
They need an ancilla
of minimal lowest dimension that we 
determined. They may be considered as templates for end-states of continuous dissipative processes that lead to quantum state synchronization, opening the path to identifying  hitherto unknown interactions and damping mechanisms that 
generate macroscopic quantum phenomena. We demonstrated QSS experimentally on an IBM quantum computer  
for the simplest set-up of two working qubits with a single ancilliary qubit,  
and introduced an asynchronicity measure capable of quantifying the 
hardware errors. 

After completion of this work we became aware of \cite{buca_algebraic_2021}, in which quantum synchronization of Markovian open quantum systems was classified based on the dynamics.
\\

\begin{acknowledgments}
We thank Roberta Zambrini, Gianluca Giorgi, and Christoph Bruder for useful communication. Financial support by Narodowe Centrum Nauki under the grant numbers DEC-2015/18/A/ST2/00274 and 2019/35/O/ST2/01049 and by Foundation 
for Polish Science under the Team-Net NTQC project is gratefully acknowledged.
We also acknowledge use of the IBM Q for this work. 
The views expressed here are those of the authors and do not reflect
the official policy or position of IBM or the IBM Q team.
\end{acknowledgments}

\onecolumngrid
\appendix

\section{Restriction to pure states}
To see that it is enough to consider pure states as input for finding
the most general quantum-state synchronizer (QSSR), first represent the state of a single subsystem in
the SU(2)-coherent state representation, 
\begin{equation}
	\label{eq:Prep}
	\rho=\int d^2\alpha P(\alpha)\ket{\alpha}\bra{\alpha}\,.  
\end{equation}
Here, the $\ket{\alpha}$ are SU(2)  coherent states labelled with the
complex parameter $\alpha$.  By stereographic projection,  we can
consider $\alpha$ represented alternatively in terms of polar and
azimuthal angle on the unit sphere. The function $P$ can then be
chosen as a real
smooth function on the sphere that exists for all states in arbitrary
finite dimensions, see
e.g.~\cite{Giraud08}. In particular, we  can expand all hermitian
$\rho_{i}^{(j)}$ operators that form a basis of $\mathcal B(\mathcal H_j)$ in
the form  \eqref{eq:Prep} with a $P$-function $P_{i_j}^{(j)}$. An arbitrary
state in $\mathcal B(\mathcal H_S)$ can then be written as 
\begin{eqnarray}
	\label{eq:rhoP}
	\rho&=&\sum_{i_1,\ldots,i_n=1}^{N^2}c_{i_1,\ldots,i_n}\rho_{i_1}^{(1)}\otimes\ldots\otimes \rho_{i_n}^{(n)}\\
	&=& \int d^{2n}\alpha P(\bm \alpha)\ket{\bm\alpha}\bra{\bm\alpha}\,,\\
	P(\bm \alpha)&\equiv& \sum_{i_1,\ldots,i_n=1}^{N^2}c_{i_1,\ldots,i_n}P_{i_1}^{(1)}(\alpha_1)\cdot\ldots\cdot P_{i_1}^{(n)}(\alpha_n)\,,          
\end{eqnarray}
with $c_{i_1,\ldots,i_n}\in \mathbb R$ and hence $P(\bm \alpha)\in \mathbb R$. 
From linearity of the partial trace it then follows immediately that
if a QSSR synchronizes all pure product states, hence $\Tr_{\bar
	i}\Phi(\ketbra{\bm\alpha})=\Tr_{\bar
	j}\Phi(\ketbra{\bm\alpha})$ for all tensor products of SU(2)
coherent states $\ket{\bm\alpha}$ and all $i,j=1,\ldots,n$, then
also $\Tr_{\bar
	i}\Phi(\rho)=\Tr_{\bar  j}\Phi(\rho)$ for arbitrary input states $\rho$ and all $i,j$.
If a quantum channels synchronizes all pure product states it
synchronizes all input states. The converse holds trivially.\\

\section{Proof of the sharpening of condition 2}
Suppose that in a set of three subsystems we have a state $\ket{\psi}$ with an even
symmetry under $P_{12}$, and odd symmetry under $P_{13}$,
\begin{align}
	P_{12}\ket{\psi} & = \ket{\psi} & P_{13}\ket{\psi} = -\ket{\psi}
\end{align}
noted as (12+) and (13-). Then, starting with the three subsystems
ordered initially as 123, we get no sign change under $P_{12}$, noted
as (213+).  Continuing a series of alternating (13-) and
(12+) swaps, we have (231-), with the antisymmetry indicated by the -
sign, (321-), (312+), (132+), (123-), i.e.~the state has changed the
sign even though the ordering of the systems is the same as at the
beginning. In other words,
\begin{align*}
	\ket{\psi} & \rightarrow P_{12}\ket{\psi} = \ket{\psi} \\
	& \rightarrow \underbrace{P_{13}P_{12}}_{P_{123}}\ket{\psi} = -\ket{\psi} \\
	& \rightarrow \hdots \\
	& \rightarrow \underbrace{\qty(P_{13}P_{12})^3}_{P_{123}^3 = \mathbb{I}} = -\ket{\psi}
\end{align*}
where it can be proven that the eigenvalues of the 3-cycle operator $P_{123}$ are $\qty{1, e^{i\frac{2\pi}{3}}, e^{i\frac{-2\pi}{3}}}$ and cubing them gives identity, forcing the cube of the operator to be the identity operator. Hence, the symmetry must be the same for all pairs $(i,j)$.  \\

\section{Generalized representation of $P_{12}$ swap on qubits} 
In the main text we have assumed that $P_{12}^2\ket{\psi}=\ket{\psi}$.  In projective Hilbert space, where all states with the same global phase are identified, one might think about generalizing this to $P_{12}^2\ket{\psi}=e^{2i\Delta}\ket{\psi}$.  A special case is the anyonic behavior given by $P_{12}\ket{\psi}=e^{i\Delta} \ket{\psi}$ with any phase $\Delta\in\mathbb R$ \cite{wilczek_magnetic_1982,wilczek_quantum_1982}.   Different computational basis states might have different phases under the action of $P_{12}$,
\begin{align}
	P_{12}\ket{00} = e^{i\Delta}\ket{00} &&
	P_{12}\ket{01} = e^{i(\Delta + \phi_{01})}\ket{10} \nonumber\\
	P_{12}\ket{10} = e^{i(\Delta + \phi_{10})}\ket{01} &&
	P_{12}\ket{11} = e^{i(\Delta + \phi_{11})}\ket{11}\,,
\end{align}
where the phase related to $\ket{00}$ defines a global phase w.r.t.~which the others are measured.  We will make two natural assumptions, however. First, for an arbitrary state $\ket{\psi} = \sum_{i,j=0}^1a_{ij}\ket{ij}$, $P_{12}^2$ should return the same state in projective Hilbert space, i.e.
\begin{align*}
	P_{12}^2\ket{\psi} = &  e^{2i\Delta}\left(a_{00}\ket{00} + e^{i(\phi_{01} + \phi_{10})}\qty(a_{01}\ket{01} + a_{10}\ket{10}) + e^{2 i \phi_{11}}a_{11}\ket{11}\right) \\
	= & e^{2
		i\Delta}\ket{\psi}\,.
\end{align*}
This is a physical request: after permuting back the state must differ at most by a global phase.  
This implies
\begin{align}
	\phi_{11} =  n \pi\,,
	\phi_{01} + \phi_{10} = 2 m \pi\,,\,\,n,m\in\mathbb{Z}\,.
\end{align}
The second one is that for ``double states'', $\ket{\psi_2} = a_{00}\ket{00} + a_{11}\ket{11}$ should stay identical up to a global phase, $P_{12}^2\ket{\psi_2} = e^{2i\Delta}\ket{\psi_2}$, as physically one cannot keep track of the number of permutations of such states even modulo 1, such that any differing phases for different double states would be ill defined. From this we get $\phi_{11} = 0$, i.e.~in the subspace $\text{span}(\ket{00},\ket{11})$, $P_{12}$ is represented by $e^{i\Delta}\mathbb{I}$, whereas 
in the subspace $\text{span}(\ket{01},\ket{10})$, $P_{12}$ is represented by 
\begin{equation}
	P_{12} = e^{i\Delta}\mqty(0 & e^{i \phi_{01}} \\ e^{-i \phi_{01}} & 0) = e^{i\Delta}(\cos(\phi_{01})\sigma_x + \sin(\phi_{01}) \sigma_y)\,.
\end{equation}
Hence, the eigenvalues of $P_{12}$ are $\pm e^{i \Delta}$ (with $e^{i\Delta}$ three-fold degenerate).  The corresponding subspaces correspond to the usual triplet and singlet states of two spins-1/2, which now aquire the global phases $\pm e^{i \Delta}$.
If we define the $\ket{s_{ij}}$ and $\ket{a_{ij}}$ as basis vectors in these two subspaces, we recover up to the global phases the situtation studied in the main text, and Proposition 1 applies accordingly.\\
Additional freedom arises, however, from the fact that the permutations act on the state in the full Hilbert space $\mathcal H$ when we act on columns of $U$.
I.e.~there is an additional label for the ancilla, and so $P_{12}$ can act differently in different subspaces labelled by different ancillary states, see the first example below. In the second example we use two different representations of $P_{12}$ that differ by the phases $\phi_{01},\phi_{10}$ and hence lead to different subspaces with eigenvalue 1 that are then mixed.\\
The considerations extend to higher dimension, yielding a single $\Delta$ global phase and $\frac{N(N-1)}{2}$, which corresponds to $\frac{1}{2}\text{dim}\qty(\text{span} \qty{\ket{ij}: i\neq j})$. Moreover, for $n$ primary subsystems one can select $n-1$ exchange operators which generate the whole permutation group $S_n$ and for each of them representation is independent. Thus, the total number of degrees of freedom provided by the SWAP operators is given by $\qty(n-1)\qty(\frac{N(N-1)}{2}+1)$.

\section{
	Connection to representations of SU(N)
}
Here we give more details on the construction of the QSSRs, eq.(9) in the main text.
As mentioned there, the fully symmetric or fully anti-symmetric QSSRs can be constructed as maximal or minimal angular momentum representations of SU(N).  The fully anti-symmetric QSSRs exist only for sufficiently large local dimensions, as can be seen by using Young tableaus. Consider the following SU(2) examples:
\begin{align}
	\ydiagram{1}\otimes\ydiagram{1} & = \underbrace{\rd{\ydiagram{2}}}_{\text{symmetric space}} \oplus \underbrace{\rd{\ydiagram{1,1}}}_{\text{antisymmetric space}} \\
	\ydiagram{1}\otimes\ydiagram{1}\otimes\ydiagram{1} & = \qty(\ydiagram{2}\oplus\ydiagram{1,1})\otimes\ydiagram{1} \nonumber\\
	& = \underbrace{\rd{\ydiagram{3}}}_{\text{symmetric space}} \oplus \ydiagram{2,1} \oplus \ydiagram{2,1}
\end{align}
and there is no antisymmetric space 
beyond 
$n=2$.

The SU(3) case is slightly different:
\begin{align}
	\ydiagram{1}\otimes\ydiagram{1} & = \rd{\ydiagram{2}} \oplus \rd{\ydiagram{1,1}} \\
	\ydiagram{1}\otimes\ydiagram{1}\otimes\ydiagram{1} & =\rd{ \ydiagram{3}} \oplus \ydiagram{2,1} \oplus \ydiagram{2,1} \oplus \underbrace{\rd{\ydiagram{1,1,1}}}_{\text{Antisymmetric space}}\,.
\end{align}
Here, an antisymmetric subspace appears, as we can antisymmetrize three different symbols.

As a final case, 
consider four parties from SU(4)
\begin{equation}
	\ydiagram{1}\otimes\ydiagram{1}\otimes\ydiagram{1}\otimes\ydiagram{1} = \rd{\ydiagram{4}} \oplus 3\,\ydiagram{3,1}\oplus2\,\ydiagram{2,2}\oplus3\,\ydiagram{2,1,1}\oplus\rd{\ydiagram{1,1,1,1}}\,.
\end{equation}
In general, the full columns and full rows 
of the Young tableaus 
generate the fully symmetric and fully antisymmetric 
representations, respectively, which 
constitute SQSs and thus a basis for QSSRs.

\section{Examples of mixed symmetry synchronizing channels}

\noindent\emph{Mixing $\pm$ symmetries between Kraus operators} --- The symmetry of subspaces to which the columns of Kraus operators are related to can be different for each operator. An example of such a QSSR is given by the following Kraus representation
\begin{align}
	K_0 & = \frac{1}{2}\mqty(
	0 & \wm0 & \wm0 & 0 \\
	0 & \wm1 & -1 & 0 \\
	0 & -1 & \wm1 & 0 \\
	0 & \wm0 & \wm0 & 0) &
	K_1 & = \frac{1}{2}\mqty(
	2 & 0 & 0 & 0 \\
	0 & 1 & 1 & 0 \\
	0 & 1 & 1 & 0 \\
	0 & 0 & 0 & 2)\,.
\end{align}
For $K_0$ the columns are antisymmetric with respect to $P_{12}$, as opposed to the symmetry in $K_1$. This case can be written in terms of projectors onto $\ket{j,m}$ states of joint angular momentum of two spin-1/2 particles,
\begin{align}
	K_0 = \op{0,0} && K_1 = \sum_{m = -1}^1 \op{1, m}\,.
\end{align}
The example can be extended to two systems with arbitrary $N$ and $M=2$
by considering $K_0$ and $K_1$ as 
projectors onto SU(N) irreps with extremal weights. \\ 

\noindent\emph{Mixing representation of exchange operator between the Kraus operators} --- Here we consider two very specific, perfectly admissible representations of the exchange operator:
\begin{align}
	P_{12} & = \mqty(1 & 0 & 0 & 0 \\ 0 & 0 & 1 & 0 \\ 0 & 1 & 0 & 0 \\  0 & 0 & 0 & 1) &
	\text{+1 eigenvectors:}&\quad \mqty{
		\mqty(1 \\ 0 \\ 0 \\ 0 ) & 
		\mqty(0 \\ 1 \\ 1 \\ 0 ) & 
		\mqty(0 \\ 0 \\ 0 \\ 1 )} & 
	\text{-1 eigenvectors:}&\quad \mqty{
		\mqty(\wm0 \\ \wm1 \\ -1 \\ \wm0 )}\\
	P'_{12} & = \mqty(1 & 0 & 0 & 0 \\ 0 & 0 & i & 0 \\ 0 & -i & 0 & 0 \\  0 & 0 & 0 & 1) &
	\text{+1 eigenvectors:}&\quad \mqty{
		\mqty(1 \\ 0 \\ 0 \\ 0 ) & 
		\mqty(\wm0 \\ \wm1 \\ -i \\ \wm0 ) & 
		\mqty(0 \\ 0 \\ 0 \\ 1 )} & 
	\text{-1 eigenvectors:}&\quad \mqty{
		\mqty(0 \\ 1 \\ i \\ 0 )}
\end{align}
We can assign $P_{12}$ to the first Kraus operator and $P'_{12}$ to the second and choose symmetric subspaces for both of them. All the steps of construction are the same as in the main body. 
In this way we produce a valid synchronization channel, 
\begin{align}
	K_1 & = \frac{1}{2}\mqty(
	2 & 0 & 0 & 0 \\ 
	0 & 1 & 1 & 0 \\
	0 & 1 & 1 & 0 \\
	0 & 0 & 0 & 2) &
	K_2 & = \frac{1}{2}\mqty(
	0 & \wm0 & \wm0 & 0 \\
	0 & \wm1 & -1 & 0 \\
	0 & -i & \wm i & 0 \\
	0 & \wm0 & \wm0 & 0)\,.
\end{align}
One can prove by rudimentary calculations that such a channel is still a QSSR.\\

\noindent\emph{Total space of QSSRs} --- In summary, for a fixed local dimension $N$, number of primary subsystems $n$ and the ancillary dimension $M$, we find at most $M+1$ disjoint manifolds, distinguished by the number of symmetric Kraus operators $M_{sym}$ and the antisymmetric ones $M_{asym} = M - M_{sym}$. The real dimensionality of each manifold is given by
$$
\underbrace{M\qty(n-1)\qty(\frac{N(N-1)}{2} + 1)}_{\text{SWAP operators' freedom}} + \underbrace{N^n\qty(2\qty(M_{sym}\binom{n+N-1}{n} + M_{asym}\binom{N}{n}) - N^n)}_{\text{Choice of $N^n$ orthonormal vectors from proper-dimensional $\mathbb{C}$ vector space}}
$$

\section{Quantum circuit}  
\noindent In order to execute the quantum channel (10) on a quantum device, the corresponding unitary must be written as a quantum circuit consisting of quantum gates. This unitary can be constructed by using only CNOT and controlled Hadamard gates. A CNOT gate acts on two qubits by swapping the states $\ket{0}$ and $\ket{1}$ of the target qubit if the control qubit is in the state $\ket{1}$. A Hadamard gate acts on one qubit by transforming $\ket{0} \rightarrow \frac{1}{\sqrt{2}}\qty( \ket{0}+\ket{1})$ and $\ket{1} \rightarrow \frac{1}{\sqrt{2}}\qty( \ket{0}-\ket{1})$. A controlled Hadamard gate is simply a Hadamard gate that acts on the target qubit only if the control qubit is in the state $\ket{1}$. The constructed circuit can be seen below in Fig. \ref{const_circ}.

\begin{figure*}[htb!]
	\includegraphics[width=0.8\textwidth, height=2.5cm,trim={0 1cm 0 1.5cm},clip]{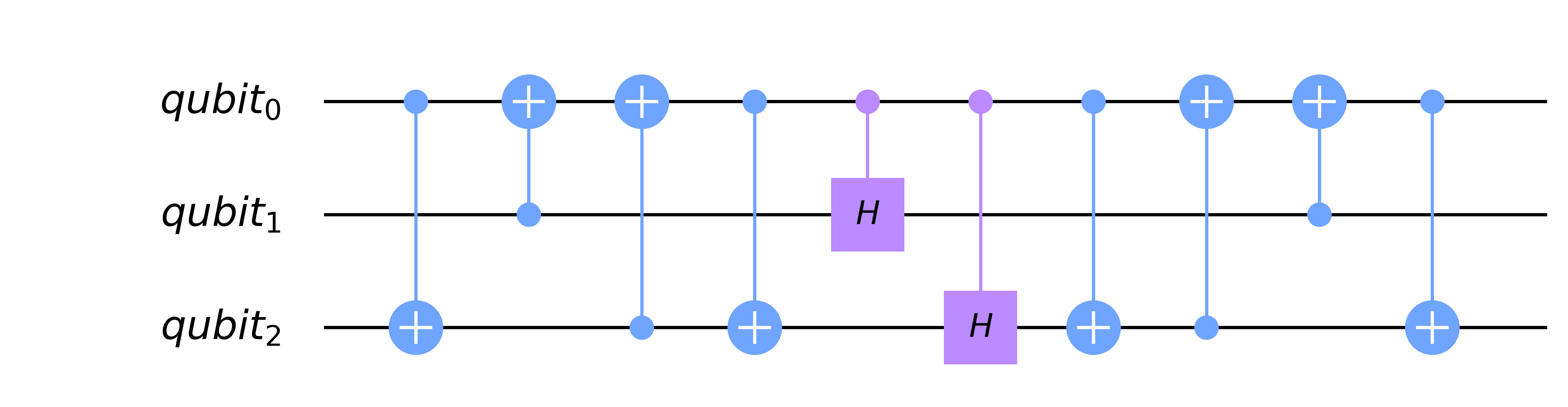}
	\caption{\label{const_circ} The constructed quantum circuit acting on 3 qubits that implements the quantum channel (10) on qubits 0 and 1, consisting only of CNOTs and controlled Hadamard gates.}
\end{figure*}

\begin{figure*}[htb!]
	\begin{center}
		\hspace*{-2.7cm}\includegraphics[width=20.5cm,trim={0 1cm 0 1.5cm},clip]{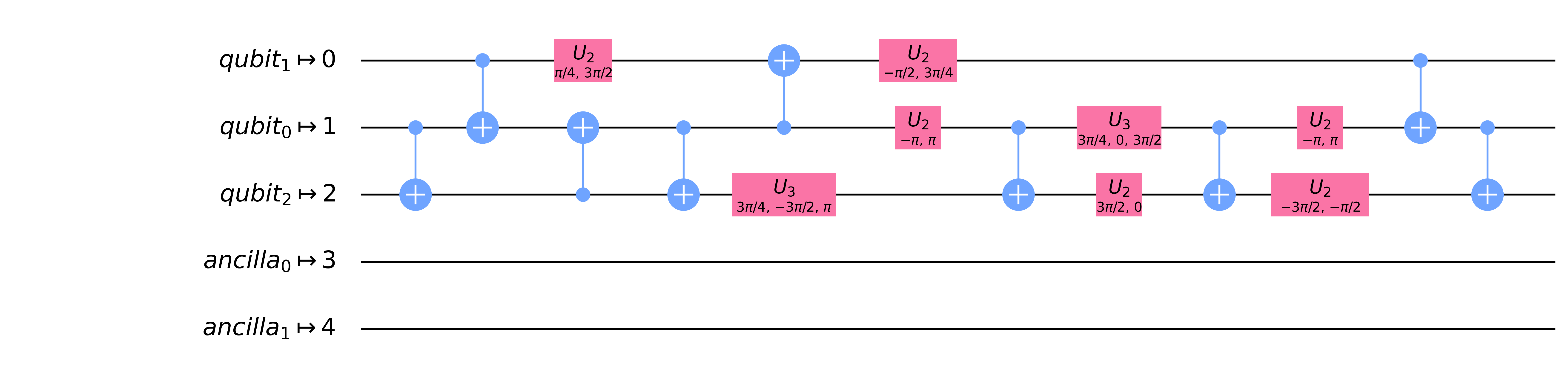} 
		\caption{\label{transpiled} Transpiled circuit of the exemplary quantum channel (10) run on the IBM Santiago. Note the changed assignment of logical qubits to physical ones. To extract the Bloch vector components, we need to append up to two more gates for measurements as well as one gate for state preparation on each wire, thus resulting in a critical path length of 15.}
	\end{center}
\end{figure*}

\noindent Unfortunately, this circuit cannot be directly implemented on the IBM Santiago quantum device but must first be transpiled. This is due to the linear qubit layout of the quantum computer which only allows for the application of two qubit gates onto direct neighbours.  Furthermore, the device has a specific set of basis gates of which the circuit needs to consist of. The constructed circuit must therefore be decomposed into those basis gates while respecting the qubit layout.  The implementable, transpiled circuit can be seen in Fig. \ref{transpiled}. The full circuit has two more gates for state preparation and up to six more gates for the measurement of the Bloch vector components. The initial states were chosen by creating a product state out of uniformly sampled pure single-qubit states with respect to the Haar measure.

\noindent The $U_3$ gate is defined as follows:
\begin{equation*}
	U_3(\theta, \phi,\lambda) =
	\begin{pmatrix}
		\cos(\theta/2) &  -e^{i\lambda} \sin(\theta/2)     \\
		e^{i\phi} \sin(\theta/2)  &  e^{i(\phi+\lambda)} \cos(\theta/2)    
	\end{pmatrix},
\end{equation*}
\noindent while $U_2(\phi, \lambda) = U_3(\pi/2, \phi, \lambda)$.\\

\begin{table}[htb!]
	
	\begin{ruledtabular}
		\begin{tabular}{cccc}
			$n$&$M$ for $N=2$&$M$ for $N=3$&$M$ for $N=4$\\
			\hline
			2&2&2&2\\
			3&2&3&4\\
			4&4&6&8\\
			5&6&12&19\\
			6&10&27&49\\
			7&16&61&137\\
			8&29&146&398\\
			9&52&358&1192\\
		\end{tabular}
	\end{ruledtabular}
	\caption{\label{tab:table4} Overview over the minimal needed dimensionality $M$ of the ancilla for a different number $n$ of subsystems and dimensionality $N$ of those subsystems.}
\end{table}

If the synchronizing quantum channels in the 2$\times  2 \times 2$ case
are
additionaly optimized with respect to creating 
maximum purity, a new property 
emerges for a
single-qubit ancilla: The ancilla qubit ends up in 
the initial state of one of the 
system qubits in analogy to the process of teleportation.  
A similar experiment on a 4-qubit system was conducted but no
conclusive results could be found due to 
the large circuit depth, incompatible with the noisy device.

\twocolumngrid

\bibliography{../../bibfile_master/mybibs_bt}

\begin{thebibliography}{30}%
\makeatletter
\providecommand \@ifxundefined [1]{%
 \@ifx{#1\undefined}
}%
\providecommand \@ifnum [1]{%
 \ifnum #1\expandafter \@firstoftwo
 \else \expandafter \@secondoftwo
 \fi
}%
\providecommand \@ifx [1]{%
 \ifx #1\expandafter \@firstoftwo
 \else \expandafter \@secondoftwo
 \fi
}%
\providecommand \natexlab [1]{#1}%
\providecommand \enquote  [1]{``#1''}%
\providecommand \bibnamefont  [1]{#1}%
\providecommand \bibfnamefont [1]{#1}%
\providecommand \citenamefont [1]{#1}%
\providecommand \href@noop [0]{\@secondoftwo}%
\providecommand \href [0]{\begingroup \@sanitize@url \@href}%
\providecommand \@href[1]{\@@startlink{#1}\@@href}%
\providecommand \@@href[1]{\endgroup#1\@@endlink}%
\providecommand \@sanitize@url [0]{\catcode `\\12\catcode `\$12\catcode
  `\&12\catcode `\#12\catcode `\^12\catcode `\_12\catcode `\%12\relax}%
\providecommand \@@startlink[1]{}%
\providecommand \@@endlink[0]{}%
\providecommand \url  [0]{\begingroup\@sanitize@url \@url }%
\providecommand \@url [1]{\endgroup\@href {#1}{\urlprefix }}%
\providecommand \urlprefix  [0]{URL }%
\providecommand \Eprint [0]{\href }%
\providecommand \doibase [0]{https://doi.org/}%
\providecommand \selectlanguage [0]{\@gobble}%
\providecommand \bibinfo  [0]{\@secondoftwo}%
\providecommand \bibfield  [0]{\@secondoftwo}%
\providecommand \translation [1]{[#1]}%
\providecommand \BibitemOpen [0]{}%
\providecommand \bibitemStop [0]{}%
\providecommand \bibitemNoStop [0]{.\EOS\space}%
\providecommand \EOS [0]{\spacefactor3000\relax}%
\providecommand \BibitemShut  [1]{\csname bibitem#1\endcsname}%
\let\auto@bib@innerbib\@empty
\bibitem [{\citenamefont {Pikovsky}\ \emph {et~al.}(2001)\citenamefont
  {Pikovsky}, \citenamefont {Rosenblum},\ and\ \citenamefont
  {Kurths}}]{PRK2001}%
  \BibitemOpen
  \bibfield  {author} {\bibinfo {author} {\bibfnamefont {A.}~\bibnamefont
  {Pikovsky}}, \bibinfo {author} {\bibfnamefont {M.}~\bibnamefont
  {Rosenblum}},\ and\ \bibinfo {author} {\bibfnamefont {J.}~\bibnamefont
  {Kurths}},\ }\href@noop {} {\emph {\bibinfo {title} {Synchronization. A
  Universal Concept in Nonlinear Sciences}}}\ (\bibinfo  {publisher} {Cambridge
  University Press, Cambridge, UK},\ \bibinfo {year} {2001})\BibitemShut
  {NoStop}%
\bibitem [{\citenamefont {Walter}\ \emph {et~al.}(2014)\citenamefont {Walter},
  \citenamefont {Nunnenkamp},\ and\ \citenamefont
  {Bruder}}]{walter_quantum_2014}%
  \BibitemOpen
  \bibfield  {author} {\bibinfo {author} {\bibfnamefont {S.}~\bibnamefont
  {Walter}}, \bibinfo {author} {\bibfnamefont {A.}~\bibnamefont {Nunnenkamp}},\
  and\ \bibinfo {author} {\bibfnamefont {C.}~\bibnamefont {Bruder}},\
  }\bibfield  {title} {\bibinfo {title} {Quantum {Synchronization} of a
  {Driven} {Self}-{Sustained} {Oscillator}},\ }\href
  {https://doi.org/10.1103/PhysRevLett.112.094102} {\bibfield  {journal}
  {\bibinfo  {journal} {Phys. Rev. Lett.}\ }\textbf {\bibinfo {volume} {112}},\
  \bibinfo {pages} {094102} (\bibinfo {year} {2014})}\BibitemShut {NoStop}%
\bibitem [{\citenamefont {Walter}\ \emph {et~al.}(2015)\citenamefont {Walter},
  \citenamefont {Nunnenkamp},\ and\ \citenamefont
  {Bruder}}]{walter_quantum_2015}%
  \BibitemOpen
  \bibfield  {author} {\bibinfo {author} {\bibfnamefont {S.}~\bibnamefont
  {Walter}}, \bibinfo {author} {\bibfnamefont {A.}~\bibnamefont {Nunnenkamp}},\
  and\ \bibinfo {author} {\bibfnamefont {C.}~\bibnamefont {Bruder}},\
  }\bibfield  {title} {\bibinfo {title} {Quantum synchronization of two {Van}
  der {Pol} oscillators},\ }\href {https://doi.org/10.1002/andp.201400144}
  {\bibfield  {journal} {\bibinfo  {journal} {Annalen der Physik}\ }\textbf
  {\bibinfo {volume} {527}},\ \bibinfo {pages} {131} (\bibinfo {year}
  {2015})}\BibitemShut {NoStop}%
\bibitem [{\citenamefont {DeVille}(2018)}]{DeVille2018SynchronizationAS}%
  \BibitemOpen
  \bibfield  {author} {\bibinfo {author} {\bibfnamefont {L.}~\bibnamefont
  {DeVille}},\ }\bibfield  {title} {\bibinfo {title} {Synchronization and
  stability for quantum {Kuramoto}},\ }\href
  {https://doi.org/10.1007/s10955-018-2168-9} {\bibfield  {journal} {\bibinfo
  {journal} {Journal of Statistical Physics}\ }\textbf {\bibinfo {volume}
  {174}},\ \bibinfo {pages} {160} (\bibinfo {year} {2018})}\BibitemShut
  {NoStop}%
\bibitem [{\citenamefont {Heinrich}\ \emph {et~al.}(2011)\citenamefont
  {Heinrich}, \citenamefont {Ludwig}, \citenamefont {Qian}, \citenamefont
  {Kubala},\ and\ \citenamefont {Marquardt}}]{heinrich_collective_2011}%
  \BibitemOpen
  \bibfield  {author} {\bibinfo {author} {\bibfnamefont {G.}~\bibnamefont
  {Heinrich}}, \bibinfo {author} {\bibfnamefont {M.}~\bibnamefont {Ludwig}},
  \bibinfo {author} {\bibfnamefont {J.}~\bibnamefont {Qian}}, \bibinfo {author}
  {\bibfnamefont {B.}~\bibnamefont {Kubala}},\ and\ \bibinfo {author}
  {\bibfnamefont {F.}~\bibnamefont {Marquardt}},\ }\bibfield  {title} {\bibinfo
  {title} {Collective dynamics in optomechanical arrays},\ }\href
  {https://doi.org/10.1103/PhysRevLett.107.043603} {\bibfield  {journal}
  {\bibinfo  {journal} {Phys. Rev. Lett.}\ }\textbf {\bibinfo {volume} {107}},\
  \bibinfo {pages} {043603} (\bibinfo {year} {2011})}\BibitemShut {NoStop}%
\bibitem [{\citenamefont {Giorgi}\ \emph {et~al.}(2012)\citenamefont {Giorgi},
  \citenamefont {Galve}, \citenamefont {Manzano}, \citenamefont {Colet},\ and\
  \citenamefont {Zambrini}}]{giorgi_quantum_2012}%
  \BibitemOpen
  \bibfield  {author} {\bibinfo {author} {\bibfnamefont {G.~L.}\ \bibnamefont
  {Giorgi}}, \bibinfo {author} {\bibfnamefont {F.}~\bibnamefont {Galve}},
  \bibinfo {author} {\bibfnamefont {G.}~\bibnamefont {Manzano}}, \bibinfo
  {author} {\bibfnamefont {P.}~\bibnamefont {Colet}},\ and\ \bibinfo {author}
  {\bibfnamefont {R.}~\bibnamefont {Zambrini}},\ }\bibfield  {title} {\bibinfo
  {title} {Quantum correlations and mutual synchronization},\ }\href
  {https://doi.org/10.1103/PhysRevA.85.052101} {\bibfield  {journal} {\bibinfo
  {journal} {Phys. Rev. A}\ }\textbf {\bibinfo {volume} {85}},\ \bibinfo
  {pages} {052101} (\bibinfo {year} {2012})}\BibitemShut {NoStop}%
\bibitem [{\citenamefont {Manzano}\ \emph {et~al.}(2013)\citenamefont
  {Manzano}, \citenamefont {Galve}, \citenamefont {Giorgi}, \citenamefont
  {Hern\'andez-Garc\'{\i}a},\ and\ \citenamefont
  {Zambrini}}]{manzano_synchronization_2013}%
  \BibitemOpen
  \bibfield  {author} {\bibinfo {author} {\bibfnamefont {G.}~\bibnamefont
  {Manzano}}, \bibinfo {author} {\bibfnamefont {F.}~\bibnamefont {Galve}},
  \bibinfo {author} {\bibfnamefont {G.~L.}\ \bibnamefont {Giorgi}}, \bibinfo
  {author} {\bibfnamefont {E.}~\bibnamefont {Hern\'andez-Garc\'{\i}a}},\ and\
  \bibinfo {author} {\bibfnamefont {R.}~\bibnamefont {Zambrini}},\ }\bibfield
  {title} {\bibinfo {title} {Synchronization, quantum correlations and
  entanglement in oscillator networks},\ }\href
  {https://doi.org/10.1038/srep01439} {\bibfield  {journal} {\bibinfo
  {journal} {Scientific Reports}\ }\textbf {\bibinfo {volume} {3}} (\bibinfo
  {year} {2013})}\BibitemShut {NoStop}%
\bibitem [{\citenamefont {Ludwig}\ and\ \citenamefont
  {Marquardt}(2013)}]{ludwig_quantum_2013}%
  \BibitemOpen
  \bibfield  {author} {\bibinfo {author} {\bibfnamefont {M.}~\bibnamefont
  {Ludwig}}\ and\ \bibinfo {author} {\bibfnamefont {F.}~\bibnamefont
  {Marquardt}},\ }\bibfield  {title} {\bibinfo {title} {Quantum many-body
  dynamics in optomechanical arrays},\ }\href
  {https://doi.org/10.1103/PhysRevLett.111.073603} {\bibfield  {journal}
  {\bibinfo  {journal} {Phys. Rev. Lett.}\ }\textbf {\bibinfo {volume} {111}},\
  \bibinfo {pages} {073603} (\bibinfo {year} {2013})}\BibitemShut {NoStop}%
\bibitem [{\citenamefont {Sonar}\ \emph {et~al.}(2018)\citenamefont {Sonar},
  \citenamefont {Hajdu\ifmmode~\check{s}\else \v{s}\fi{}ek}, \citenamefont
  {Mukherjee}, \citenamefont {Fazio}, \citenamefont {Vedral}, \citenamefont
  {Vinjanampathy},\ and\ \citenamefont {Kwek}}]{PhysRevLett.120.163601}%
  \BibitemOpen
  \bibfield  {author} {\bibinfo {author} {\bibfnamefont {S.}~\bibnamefont
  {Sonar}}, \bibinfo {author} {\bibfnamefont {M.}~\bibnamefont
  {Hajdu\ifmmode~\check{s}\else \v{s}\fi{}ek}}, \bibinfo {author}
  {\bibfnamefont {M.}~\bibnamefont {Mukherjee}}, \bibinfo {author}
  {\bibfnamefont {R.}~\bibnamefont {Fazio}}, \bibinfo {author} {\bibfnamefont
  {V.}~\bibnamefont {Vedral}}, \bibinfo {author} {\bibfnamefont
  {S.}~\bibnamefont {Vinjanampathy}},\ and\ \bibinfo {author} {\bibfnamefont
  {L.-C.}\ \bibnamefont {Kwek}},\ }\bibfield  {title} {\bibinfo {title}
  {Squeezing enhances quantum synchronization},\ }\href
  {https://doi.org/10.1103/PhysRevLett.120.163601} {\bibfield  {journal}
  {\bibinfo  {journal} {Phys. Rev. Lett.}\ }\textbf {\bibinfo {volume} {120}},\
  \bibinfo {pages} {163601} (\bibinfo {year} {2018})}\BibitemShut {NoStop}%
\bibitem [{\citenamefont {Kato}\ \emph {et~al.}(2019)\citenamefont {Kato},
  \citenamefont {Yamamoto},\ and\ \citenamefont
  {Nakao}}]{PhysRevResearch.1.033012}%
  \BibitemOpen
  \bibfield  {author} {\bibinfo {author} {\bibfnamefont {Y.}~\bibnamefont
  {Kato}}, \bibinfo {author} {\bibfnamefont {N.}~\bibnamefont {Yamamoto}},\
  and\ \bibinfo {author} {\bibfnamefont {H.}~\bibnamefont {Nakao}},\ }\bibfield
   {title} {\bibinfo {title} {Semiclassical phase reduction theory for quantum
  synchronization},\ }\href {https://doi.org/10.1103/PhysRevResearch.1.033012}
  {\bibfield  {journal} {\bibinfo  {journal} {Phys. Rev. Research}\ }\textbf
  {\bibinfo {volume} {1}},\ \bibinfo {pages} {033012} (\bibinfo {year}
  {2019})}\BibitemShut {NoStop}%
\bibitem [{\citenamefont {Giorgi}\ \emph {et~al.}(2013)\citenamefont {Giorgi},
  \citenamefont {Plastina}, \citenamefont {Francica},\ and\ \citenamefont
  {Zambrini}}]{giorgi_spontaneous_2013}%
  \BibitemOpen
  \bibfield  {author} {\bibinfo {author} {\bibfnamefont {G.~L.}\ \bibnamefont
  {Giorgi}}, \bibinfo {author} {\bibfnamefont {F.}~\bibnamefont {Plastina}},
  \bibinfo {author} {\bibfnamefont {G.}~\bibnamefont {Francica}},\ and\
  \bibinfo {author} {\bibfnamefont {R.}~\bibnamefont {Zambrini}},\ }\bibfield
  {title} {\bibinfo {title} {Spontaneous synchronization and quantum
  correlation dynamics of open spin systems},\ }\href
  {https://doi.org/10.1103/PhysRevA.88.042115} {\bibfield  {journal} {\bibinfo
  {journal} {Phys. Rev. A}\ }\textbf {\bibinfo {volume} {88}},\ \bibinfo
  {pages} {042115} (\bibinfo {year} {2013})}\BibitemShut {NoStop}%
\bibitem [{\citenamefont {Zhirov}\ and\ \citenamefont
  {Shepelyansky}(2008)}]{zhirov_synchronization_2008}%
  \BibitemOpen
  \bibfield  {author} {\bibinfo {author} {\bibfnamefont {O.~V.}\ \bibnamefont
  {Zhirov}}\ and\ \bibinfo {author} {\bibfnamefont {D.~L.}\ \bibnamefont
  {Shepelyansky}},\ }\bibfield  {title} {\bibinfo {title} {Synchronization and
  {Bistability} of a {Qubit} {Coupled} to a {Driven} {Dissipative}
  {Oscillator}},\ }\href {https://doi.org/10.1103/PhysRevLett.100.014101}
  {\bibfield  {journal} {\bibinfo  {journal} {Phys. Rev. Lett.}\ }\textbf
  {\bibinfo {volume} {100}},\ \bibinfo {pages} {014101} (\bibinfo {year}
  {2008})}\BibitemShut {NoStop}%
\bibitem [{\citenamefont {Zhirov}\ and\ \citenamefont
  {Shepelyansky}(2009)}]{zhirov_quantum_2009}%
  \BibitemOpen
  \bibfield  {author} {\bibinfo {author} {\bibfnamefont {O.~V.}\ \bibnamefont
  {Zhirov}}\ and\ \bibinfo {author} {\bibfnamefont {D.~L.}\ \bibnamefont
  {Shepelyansky}},\ }\bibfield  {title} {\bibinfo {title} {Quantum
  synchronization and entanglement of two qubits coupled to a driven
  dissipative resonator},\ }\href {https://doi.org/10.1103/PhysRevB.80.014519}
  {\bibfield  {journal} {\bibinfo  {journal} {Phys. Rev. B}\ }\textbf {\bibinfo
  {volume} {80}},\ \bibinfo {pages} {014519} (\bibinfo {year}
  {2009})}\BibitemShut {NoStop}%
\bibitem [{\citenamefont {Cattaneo}\ \emph {et~al.}(2020)\citenamefont
  {Cattaneo}, \citenamefont {Giorgi}, \citenamefont {Maniscalco}, \citenamefont
  {Paraoanu},\ and\ \citenamefont {Zambrini}}]{cattaneo_synchronization_2020}%
  \BibitemOpen
  \bibfield  {author} {\bibinfo {author} {\bibfnamefont {M.}~\bibnamefont
  {Cattaneo}}, \bibinfo {author} {\bibfnamefont {G.~L.}\ \bibnamefont
  {Giorgi}}, \bibinfo {author} {\bibfnamefont {S.}~\bibnamefont {Maniscalco}},
  \bibinfo {author} {\bibfnamefont {G.~S.}\ \bibnamefont {Paraoanu}},\ and\
  \bibinfo {author} {\bibfnamefont {R.}~\bibnamefont {Zambrini}},\ }\bibfield
  {title} {\bibinfo {title} {Synchronization and subradiance as signatures of
  entangling bath between superconducting qubits},\ }\href
  {http://arxiv.org/abs/2005.06229} {\bibfield  {journal} {\bibinfo  {journal}
  {arXiv:2005.06229}\ } (\bibinfo {year} {2020})}\BibitemShut {NoStop}%
\bibitem [{\citenamefont {Koppenh\"ofer}\ \emph {et~al.}(2020)\citenamefont
  {Koppenh\"ofer}, \citenamefont {Bruder},\ and\ \citenamefont
  {Roulet}}]{PhysRevResearch.2.023026}%
  \BibitemOpen
  \bibfield  {author} {\bibinfo {author} {\bibfnamefont {M.}~\bibnamefont
  {Koppenh\"ofer}}, \bibinfo {author} {\bibfnamefont {C.}~\bibnamefont
  {Bruder}},\ and\ \bibinfo {author} {\bibfnamefont {A.}~\bibnamefont
  {Roulet}},\ }\bibfield  {title} {\bibinfo {title} {Quantum synchronization on
  the {IBM} {Q} system},\ }\href
  {https://doi.org/10.1103/PhysRevResearch.2.023026} {\bibfield  {journal}
  {\bibinfo  {journal} {Phys. Rev. Research}\ }\textbf {\bibinfo {volume}
  {2}},\ \bibinfo {pages} {023026} (\bibinfo {year} {2020})}\BibitemShut
  {NoStop}%
\bibitem [{\citenamefont {Roulet}\ and\ \citenamefont
  {Bruder}(2018{\natexlab{a}})}]{roulet_quantum_2018}%
  \BibitemOpen
  \bibfield  {author} {\bibinfo {author} {\bibfnamefont {A.}~\bibnamefont
  {Roulet}}\ and\ \bibinfo {author} {\bibfnamefont {C.}~\bibnamefont
  {Bruder}},\ }\bibfield  {title} {\bibinfo {title} {Quantum synchronization
  and entanglement generation},\ }\href
  {https://doi.org/10.1103/PhysRevLett.121.063601} {\bibfield  {journal}
  {\bibinfo  {journal} {Phys. Rev. Lett.}\ }\textbf {\bibinfo {volume} {121}},\
  \bibinfo {pages} {063601} (\bibinfo {year} {2018}{\natexlab{a}})}\BibitemShut
  {NoStop}%
\bibitem [{\citenamefont {Fiderer}\ \emph {et~al.}(2016)\citenamefont
  {Fiderer}, \citenamefont {Ku{\'s}},\ and\ \citenamefont
  {Braun}}]{fiderer_quantum-phase_2016}%
  \BibitemOpen
  \bibfield  {author} {\bibinfo {author} {\bibfnamefont {L.~J.}\ \bibnamefont
  {Fiderer}}, \bibinfo {author} {\bibfnamefont {M.}~\bibnamefont {Ku{\'s}}},\
  and\ \bibinfo {author} {\bibfnamefont {D.}~\bibnamefont {Braun}},\ }\bibfield
   {title} {\bibinfo {title} {Quantum-phase synchronization},\ }\href
  {https://doi.org/10.1103/PhysRevA.94.032336} {\bibfield  {journal} {\bibinfo
  {journal} {Phys. Rev. A}\ }\textbf {\bibinfo {volume} {94}},\ \bibinfo
  {pages} {032336} (\bibinfo {year} {2016})}\BibitemShut {NoStop}%
\bibitem [{\citenamefont {Jaseem}\ \emph {et~al.}(2020)\citenamefont {Jaseem},
  \citenamefont {Hajdu{\v s}ek}, \citenamefont {Vedral}, \citenamefont {Fazio},
  \citenamefont {Kwek},\ and\ \citenamefont
  {Vinjanampathy}}]{Jaseem2020QuantumSI}%
  \BibitemOpen
  \bibfield  {author} {\bibinfo {author} {\bibfnamefont {N.}~\bibnamefont
  {Jaseem}}, \bibinfo {author} {\bibfnamefont {M.}~\bibnamefont {Hajdu{\v
  s}ek}}, \bibinfo {author} {\bibfnamefont {V.}~\bibnamefont {Vedral}},
  \bibinfo {author} {\bibfnamefont {R.}~\bibnamefont {Fazio}}, \bibinfo
  {author} {\bibfnamefont {L.-C.}\ \bibnamefont {Kwek}},\ and\ \bibinfo
  {author} {\bibfnamefont {S.}~\bibnamefont {Vinjanampathy}},\ }\bibfield
  {title} {\bibinfo {title} {Quantum synchronization in nanoscale heat
  engines.},\ }\href {https://doi.org/10.1103/PhysRevE.101.020201} {\bibfield
  {journal} {\bibinfo  {journal} {Phys. Rev. E}\ }\textbf {\bibinfo {volume}
  {101}},\ \bibinfo {pages} {020201} (\bibinfo {year} {2020})}\BibitemShut
  {NoStop}%
\bibitem [{\citenamefont {Wootters}\ and\ \citenamefont
  {Zurek}(1982)}]{Wootters82}%
  \BibitemOpen
  \bibfield  {author} {\bibinfo {author} {\bibfnamefont {W.~K.}\ \bibnamefont
  {Wootters}}\ and\ \bibinfo {author} {\bibfnamefont {W.~H.}\ \bibnamefont
  {Zurek}},\ }\bibfield  {title} {\bibinfo {title} {A single quantum state
  cannot be cloned},\ }\href {https://doi.org/10.1038/299802a0} {\bibfield
  {journal} {\bibinfo  {journal} {Nature}\ }\textbf {\bibinfo {volume} {299}},\
  \bibinfo {pages} {802} (\bibinfo {year} {1982})}\BibitemShut {NoStop}%
\bibitem [{\citenamefont {Gisin}\ and\ \citenamefont
  {Massar}(1997)}]{GisinM97}%
  \BibitemOpen
  \bibfield  {author} {\bibinfo {author} {\bibfnamefont {N.}~\bibnamefont
  {Gisin}}\ and\ \bibinfo {author} {\bibfnamefont {S.}~\bibnamefont {Massar}},\
  }\bibfield  {title} {\bibinfo {title} {Optimal quantum cloning machines},\
  }\href {https://doi.org/10.1103/PhysRevLett.79.2153} {\bibfield  {journal}
  {\bibinfo  {journal} {Phys. Rev. Lett.}\ }\textbf {\bibinfo {volume} {79}},\
  \bibinfo {pages} {2153} (\bibinfo {year} {1997})}\BibitemShut {NoStop}%
\bibitem [{\citenamefont {Barnum}\ \emph {et~al.}(1996)\citenamefont {Barnum},
  \citenamefont {Caves}, \citenamefont {Fuchs}, \citenamefont {Jozsa},\ and\
  \citenamefont {Schumacher}}]{Barnum96}%
  \BibitemOpen
  \bibfield  {author} {\bibinfo {author} {\bibfnamefont {H.}~\bibnamefont
  {Barnum}}, \bibinfo {author} {\bibfnamefont {C.~M.}\ \bibnamefont {Caves}},
  \bibinfo {author} {\bibfnamefont {C.~A.}\ \bibnamefont {Fuchs}}, \bibinfo
  {author} {\bibfnamefont {R.}~\bibnamefont {Jozsa}},\ and\ \bibinfo {author}
  {\bibfnamefont {B.}~\bibnamefont {Schumacher}},\ }\bibfield  {title}
  {\bibinfo {title} {Noncommuting mixed states cannot be broadcast},\ }\href
  {https://doi.org/10.1103/PhysRevLett.76.2818} {\bibfield  {journal} {\bibinfo
   {journal} {Phys. Rev. Lett.}\ }\textbf {\bibinfo {volume} {76}},\ \bibinfo
  {pages} {2818} (\bibinfo {year} {1996})}\BibitemShut {NoStop}%
\bibitem [{\citenamefont {Smaczy{\'n}ski}\ \emph {et~al.}(2016)\citenamefont
  {Smaczy{\'n}ski}, \citenamefont {Roga},\ and\ \citenamefont
  {{\.Z}yczkowski}}]{SRZ16}%
  \BibitemOpen
  \bibfield  {author} {\bibinfo {author} {\bibfnamefont {M.}~\bibnamefont
  {Smaczy{\'n}ski}}, \bibinfo {author} {\bibfnamefont {W.}~\bibnamefont
  {Roga}},\ and\ \bibinfo {author} {\bibfnamefont {K.}~\bibnamefont
  {{\.Z}yczkowski}},\ }\bibfield  {title} {\bibinfo {title} {Selfcomplementary
  quantum channels},\ }\href {https://doi.org/10.1142/S1230161216500141}
  {\bibfield  {journal} {\bibinfo  {journal} {Open Systems Inform. Dynamics}\
  }\textbf {\bibinfo {volume} {23}} (\bibinfo {year} {2016})}\BibitemShut
  {NoStop}%
\bibitem [{\citenamefont {Czartowski}\ \emph {et~al.}(2019)\citenamefont
  {Czartowski}, \citenamefont {Braun},\ and\ \citenamefont
  {Życzkowski}}]{czartowski_trade-off_2019}%
  \BibitemOpen
  \bibfield  {author} {\bibinfo {author} {\bibfnamefont {J.}~\bibnamefont
  {Czartowski}}, \bibinfo {author} {\bibfnamefont {D.}~\bibnamefont {Braun}},\
  and\ \bibinfo {author} {\bibfnamefont {K.}~\bibnamefont {Życzkowski}},\
  }\bibfield  {title} {\bibinfo {title} {Trade-off relations for operation
  entropy of complementary quantum channels},\ }\href
  {https://doi.org/10.1142/S0219749919500461} {\bibfield  {journal} {\bibinfo
  {journal} {Int. J. Quantum Inf.}\ }\textbf {\bibinfo {volume} {17}},\
  \bibinfo {pages} {1950046} (\bibinfo {year} {2019})}\BibitemShut {NoStop}%
\bibitem [{\citenamefont {Roulet}\ and\ \citenamefont
  {Bruder}(2018{\natexlab{b}})}]{PhysRevLett.121.053601}%
  \BibitemOpen
  \bibfield  {author} {\bibinfo {author} {\bibfnamefont {A.}~\bibnamefont
  {Roulet}}\ and\ \bibinfo {author} {\bibfnamefont {C.}~\bibnamefont
  {Bruder}},\ }\bibfield  {title} {\bibinfo {title} {Synchronizing the smallest
  possible system},\ }\href {https://doi.org/10.1103/PhysRevLett.121.053601}
  {\bibfield  {journal} {\bibinfo  {journal} {Phys. Rev. Lett.}\ }\textbf
  {\bibinfo {volume} {121}},\ \bibinfo {pages} {053601} (\bibinfo {year}
  {2018}{\natexlab{b}})}\BibitemShut {NoStop}%
\bibitem [{\citenamefont {Bacon}\ \emph {et~al.}(2006)\citenamefont {Bacon},
  \citenamefont {Chuang},\ and\ \citenamefont
  {Harrow}}]{PhysRevLett.97.170502}%
  \BibitemOpen
  \bibfield  {author} {\bibinfo {author} {\bibfnamefont {D.}~\bibnamefont
  {Bacon}}, \bibinfo {author} {\bibfnamefont {I.~L.}\ \bibnamefont {Chuang}},\
  and\ \bibinfo {author} {\bibfnamefont {A.~W.}\ \bibnamefont {Harrow}},\
  }\bibfield  {title} {\bibinfo {title} {Efficient quantum circuits for {Schur}
  and {Clebsch-Gordan} transforms},\ }\href
  {https://doi.org/10.1103/PhysRevLett.97.170502} {\bibfield  {journal}
  {\bibinfo  {journal} {Phys. Rev. Lett.}\ }\textbf {\bibinfo {volume} {97}},\
  \bibinfo {pages} {170502} (\bibinfo {year} {2006})}\BibitemShut {NoStop}%
\bibitem [{\citenamefont {Satya~Sainadh}(2013)}]{u_efficient_2013}%
  \BibitemOpen
  \bibfield  {author} {\bibinfo {author} {\bibfnamefont {U.}~\bibnamefont
  {Satya~Sainadh}},\ }\bibfield  {title} {\bibinfo {title} {An {Efficient}
  {Quantum} {Algorithm} and {Circuit} to {Generate} {Eigenstates} of {SU}(2)
  and {SU}(3) {Representations}},\ }\href {http://arxiv.org/abs/1309.2736}
  {\bibfield  {journal} {\bibinfo  {journal} {arXiv:1309.2736}\ } (\bibinfo
  {year} {2013})}\BibitemShut {NoStop}%
\bibitem [{\citenamefont {Buca}\ \emph {et~al.}(2021)\citenamefont {Buca},
  \citenamefont {Booker},\ and\ \citenamefont {Jaksch}}]{buca_algebraic_2021}%
  \BibitemOpen
  \bibfield  {author} {\bibinfo {author} {\bibfnamefont {B.}~\bibnamefont
  {Buca}}, \bibinfo {author} {\bibfnamefont {C.}~\bibnamefont {Booker}},\ and\
  \bibinfo {author} {\bibfnamefont {D.}~\bibnamefont {Jaksch}},\ }\bibfield
  {title} {\bibinfo {title} {Algebraic {Theory} of {Quantum} {Synchronization}
  and {Limit} {Cycles} under {Dissipation}},\ }\href
  {http://arxiv.org/abs/2103.01808} {\bibfield  {journal} {\bibinfo  {journal}
  {arXiv:2103.01808}\ } (\bibinfo {year} {2021})}\BibitemShut {NoStop}%
\bibitem [{\citenamefont {Giraud}\ \emph {et~al.}(2008)\citenamefont {Giraud},
  \citenamefont {Braun},\ and\ \citenamefont {Braun}}]{Giraud08}%
  \BibitemOpen
  \bibfield  {author} {\bibinfo {author} {\bibfnamefont {O.}~\bibnamefont
  {Giraud}}, \bibinfo {author} {\bibfnamefont {P.}~\bibnamefont {Braun}},\ and\
  \bibinfo {author} {\bibfnamefont {D.}~\bibnamefont {Braun}},\ }\bibfield
  {title} {\bibinfo {title} {Classicality of spin states},\ }\href
  {https://doi.org/10.1103/PhysRevA.78.042112} {\bibfield  {journal} {\bibinfo
  {journal} {Phys. Rev. A}\ }\textbf {\bibinfo {volume} {78}},\ \bibinfo {eid}
  {042112} (\bibinfo {year} {2008})}\BibitemShut {NoStop}%
\bibitem [{\citenamefont
  {Wilczek}(1982{\natexlab{a}})}]{wilczek_magnetic_1982}%
  \BibitemOpen
  \bibfield  {author} {\bibinfo {author} {\bibfnamefont {F.}~\bibnamefont
  {Wilczek}},\ }\bibfield  {title} {\bibinfo {title} {Magnetic flux, angular
  momentum, and statistics},\ }\href
  {https://doi.org/10.1103/PhysRevLett.48.1144} {\bibfield  {journal} {\bibinfo
   {journal} {Phys. Rev. Lett.}\ }\textbf {\bibinfo {volume} {48}},\ \bibinfo
  {pages} {1144} (\bibinfo {year} {1982}{\natexlab{a}})}\BibitemShut {NoStop}%
\bibitem [{\citenamefont {Wilczek}(1982{\natexlab{b}})}]{wilczek_quantum_1982}%
  \BibitemOpen
  \bibfield  {author} {\bibinfo {author} {\bibfnamefont {F.}~\bibnamefont
  {Wilczek}},\ }\bibfield  {title} {\bibinfo {title} {Quantum mechanics of
  fractional-spin particles},\ }\href
  {https://doi.org/10.1103/PhysRevLett.49.957} {\bibfield  {journal} {\bibinfo
  {journal} {Phys. Rev. Lett.}\ }\textbf {\bibinfo {volume} {49}},\ \bibinfo
  {pages} {957} (\bibinfo {year} {1982}{\natexlab{b}})}\BibitemShut {NoStop}%
\end{thebibliography}%
\end{document}